\UseRawInputEncoding
\documentclass[3p, twocolumn]{elsarticle}

\usepackage{lineno,hyperref}
\usepackage[UKenglish]{babel}
\usepackage{enumitem} 
\usepackage{multirow,booktabs,colortbl,tabularx}
\usepackage{textcomp}
\usepackage{float}
\usepackage{adjustbox}
\usepackage{blindtext}
\usepackage{amsmath}
\usepackage{txfonts}
\usepackage{hyperref}
\usepackage{chemfig}
\usepackage{tikzpagenodes}
\usepackage{siunitx}
\usetikzlibrary{automata,positioning}
\usetikzlibrary{positioning,calc}
\usepackage{subcaption}
\usepackage{tabularx,booktabs}
\newcolumntype{Y}{>{\centering\arraybackslash}X}

\journal{Physica Medica European Journal of Medical Physics}

\usepackage{fancyhdr}

\usepackage{comment}

\makeatletter
\tikzset{
    eqn-label/.style={
      append after command = {%
        \bgroup
          [current point is local=true]
          \pgfextra{\let\tikz@save@last@fig@name=\tikz@last@fig@name\tikz@node@is@a@labelfalse
            \pgfpointanchor{current page text area}{east}
            \pgf@xa=\pgf@x
            \pgfpointanchor{\tikz@last@fig@name}{center}
            \pgf@ya=\pgf@y
          }
          node [every label,
                left
                ] at (\pgf@xa,\pgf@ya) {\refstepcounter{equation}\label{#1}\hypertarget{#1}{(\@currentlabel)}\!\!}
          \pgfextra{\global\let\tikz@last@fig@name=\tikz@save@last@fig@name}
        \egroup}
    }
}
\makeatother

\begin{document}

\begin{frontmatter}

\title{On the Eptihermal Neutron Energy Limit for Accelerator-Based Boron Neutron Capture Therapy (AB-BNCT): Study and Impact of New Energy Limits}

\author[mymainaddress]{Marine Herv\'e \corref{a}}
\author[mymainaddress]{Nadine Sauzet}
\author[mymainaddress]{Daniel Santos}

\address[mymainaddress]{Univ. Grenoble Alpes, CNRS, Grenoble INP, LPSC-IN2P3, 38000 Grenoble, France}

\cortext[a]{Corresponding author : marine.herve@lpsc.in2p3.fr}

\begin{abstract}
\begin{small}
Background and purpose: Accelerator-Based Boron Neutron Capture Therapy is a radiotherapy based on compact accelerator neutron sources requiring an epithermal neutron field for tumour irradiations. Neutrons of 10 keV are considered as the maximum optimised  energy to treat deep-seated tumours. We investigated, by means of Monte Carlo simulations, the epithermal range from 10 eV to 10 keV in order to optimise the maximum epithermal neutron energy as a function of the tumour depth.

Methods: A Snyder head phantom was simulated and mono-energetic neutrons with 4 different incident energies were used: 10 eV, 100 eV, 1 keV and 10 keV. $^{10}$B capture rates and absorbed dose composition on every tissue were calculated to describe and compare the effects of lowering the maximum epithermal energy. The Therapeutic Gain (TG) was estimated considering the whole brain volume.

Results: For tumours seated at 4 cm depth, 10 eV, 100 eV and 1 keV neutrons provided respectively 54 $\%$, 36 $\%$ and 18 $\%$ increase on the TG compared to 10 keV neutrons. Neutrons with energies between 10 eV and 1 keV provided higher TG than 10 keV neutrons for tumours seated up to 6.4 cm depth inside the head. The size of the tumour does not change these results.

Conclusions: Using lower epithermal energy neutrons for AB-BNCT tumour irradiation could improve treatment efficacy, delivering more therapeutic dose while reducing the dose in healthy tissues. This could lead to new Beam Shape Assembly designs in order to optimise the BNCT irradiation.

\end{small}
\end{abstract}


\end{frontmatter}

\section{Introduction}

Boron Neutron Capture Therapy (BNCT) is a binary radiotherapy intended for the treatment of invasive and extended tumours with complex spatial distribution or radio-chemotherapyresistants. This radiotherapy consists in fixing $^{10}$B on cancer cells through targeting carriers and then submitting the patient to an adapted epithermal energy (between 0.5 eV and 10 keV) neutron field \citep{Slatkin1991,suzuki2020}. Neutron capture by the stable isotope $^{10}$B results in the production of an alpha ($^{4}$He) particle and a lithium ($^{7}$Li) nucleus, as described in (1a) and (1b).

\begin{table*}[h!]
\begin{center}
\begin{subequations}
\label{eq1}
\noindent
\begin{tikzpicture}[node distance=0cm and 1cm, remember picture]
\node (A)
    {$^{10}B + n \longrightarrow [^{11}B*]$};
\node[above right=of A, eqn-label=eq1a] (B)
    { $^{4}He + ^{7}Li \qquad Q=2.792 MeV$};
\node[below right=of A, eqn-label=eq1b] (C)
    { $^{4}He + ^{7}Li \qquad Q=2.310 MeV \qquad + \gamma [0.48 MeV]$};
    \draw[-stealth] (A) -- ( $ (A.0)!0.3!(B.west|-A.0) $ ) |- (B.west) node[auto,pos=0.7] {${\scriptstyle 6\%}$};
    \draw[-stealth] (A) -- ( $ (A.0)!0.3!(C.west|-A.0) $ ) |- (C.west) node[auto,pos=0.7] {${\scriptstyle 94\%}$};
\end{tikzpicture}
\end{subequations}
\end{center}
\label{eq:capture}
\end{table*}

Those nuclear particles have path lengths between 4.8 and 8.9 $\mu$m in tissues and a high linear energy transfer (from $160$ to $200$ keV/$\mu$m) \citep{NCT2}. As the average cell diameter reaches $10$ $\mu$m for human tissues, neutron capture products only affect the targeted cell by releasing most of their energy inside the cell volume and sparing surrounding cells. 

As neutron capture reaction on $^{10}$B cross section increases exponentially at low neutron energies, thermal neutrons are required at the tumour level to optimise the number of neutron captures and thus the treatment efficacy. Thermal neutron fields are considered mainly for superficial cancer cases, such as superficial melanomas\citep{Pazirandeh2011}. Since neutrons are thermalised and absorbed in tissues, incident neutrons with energies higher than thermal energies are required for non-superficial tumours treatment.
However, neutrons toxicity in tissues increases significantly over $10$ keV \citep{ICRP2010}, mainly due to the increase of energy transferred by elastic scattering on hydrogen nuclei. So the usual maximum neutron energy considered is about $10$ keV, allowing to reach deep seated tumours and to reduce secondary dose which represents the dose deposited in healthy tissues \citep{Barth1992}.

The first BNCT clinical treatments were performed exclusively using nuclear research reactors  to produce neutron fields \citep{Auterinen2004,moss2004procedural,auterinen2001metamorphosis,wang2011bnct}. In order to provide facilities suitable for hospital environment, an active research is made around compact accelerator neutron sources \citep{CANS2017}. The use of low energy particle compact accelerators and targets adapted to cope with high power (15 to 30 kW) for neutron production led to the development of Accelerator-Based BNCT (AB-BCNT) facilities \citep{Kreiner2016}. In this context, an AB-BNCT facility project was launched at the Laboratory of Subatomic Physics and Cosmology (LPSC) in France \citep{delorme2016theoretical, muraz2020development}. The purpose is to use a compact accelerator coupled to a beryllium target using the $^{9}$Be(d($1.45$MeV),n)$^{10}$B reaction as a neutron source \citep{Capoulat2019}. 
In the frame of this project, one important issue is to maximise the treatment efficacy, optimising the neutron energy range used for irradiation to minimise the absorbed dose in healthy tissues during the treatment.
This article presents the study made on the effects on treatment efficacy of different neutron energy by Monte Carlo simulations. 
Its principle is based on the study of the decomposition of the dose in tumour and healthy tissues, resulting from different mono-energetic epithermal neutrons. 

First of all, the simulation setup used for this study is described. The results obtained are then discussed and the last part concludes on the optimised neutron energies proposed at the output of the beam shape assembly (BSA). All the results presented here are made in the frame of a brain cancer study but might be extended to other body organs.

\section{Material and Methods}
\subsection{Neutron interactions in tissues}
\label{sec:neutron_interactions}
$^{1}$H, $^{16}$O, $^{12}$C and $^{14}$N represent more than 99$\%$ in weight of all atoms in human brain tissues \citep{ICRU46}. The absorbed dose in healthy tissues results mainly from neutron scattering reactions and neutron capture reactions on the nuclei constitutive of the brain. As thermalisation through tissues is done by neutron scattering, it contributes to the 
secondary dose on healthy tissues. In BNCT, the total absorbed dose in tissues can be expressed as following \citep{NCT}:
\begin{equation}
D_T = D_B + D_n + D_p + D_\gamma
\label{doseeq}
\end{equation}

The biological dose uses Relative Biological Effectiveness (RBE) factors to take into account the different  biological effectiveness of each component, it can be expressed as following \citep{NCT}:
\begin{equation}
D_w = w_c * D_B + w_p * D_n + w_n * D_p + w_\gamma * D_\gamma
\label{doseeqRBE}
\end{equation}

The biological dose will be expressed in Gy[RBE] units to distinguish it from the absorbed dose expressed in Gy.
The method used to calculate the biological dose in BNCT is still an active field of research in the BNCT community \citep{gonzalez2012photon,sato2019depth}. As the aim of this study is to compare different neutron energies, average values of RBE factors are used and summarised in Table \ref{RBE_factors}. Those values are obtained considering Boronophenylalanine (BPA) as boron carrier \citep{coderre1994neutron}.

\begin{table}[H]
\begin{center}
\begin{tabular}{c||c|c|c|c} 
 \hline
 tissue & tumour & brain & skull & skin \\ 
 \hline
 $w_{c}$ & 3.8 & 1.3 & 1.3 & 2.5 \\ 
 \hline
 $w_{n}$ & 3.2 & 3.2 & 3.2 & 3.2 \\ 
 \hline
 $w_{p}$ & 3.2 & 3.2 & 3.2 & 3.2 \\ 
 \hline
 $w_{\gamma}$ & 1 & 1 & 1 & 1 \\ 
 \hline
\end{tabular}
\caption{RBE factors for BPA boron carrier}
\label{RBE_factors}
\end{center}
\end{table}

Each dose component of the total dose is associated to a different nuclear reaction, detailed below:
\begin{itemize}
\item $^{10}$B(n,$\alpha$)$^{7}$Li reaction to the called boron dose D$_{B}$, considering the ionisation energy released by the alpha particle and lithium nucleus,
\item $^{1}$H(n,n')$^{1}$H reaction to the called neutron dose D$_{n}$, considering the ionisation energy released by the hydrogen resulting from the elastic neutron scattering,
\item $^{14}$N(n,p)$^{14}$C to the called proton dose D$_{p}$, considering the ionisation energy released by the proton and the carbon nucleus,
\item $^{1}$H(n,$\gamma$)$^{2}$H reaction and the residual gamma from $^{10}$B capture reaction ($^{10}$B(n,$\gamma$)$^{7}$Li)  to the called gamma dose D$_{\gamma}$, only considering the energy transferred by the photon by pair production, Compton and photoelectric scattering.
\end{itemize}

\begin{figure*}
\centering
\includegraphics[scale=0.45]{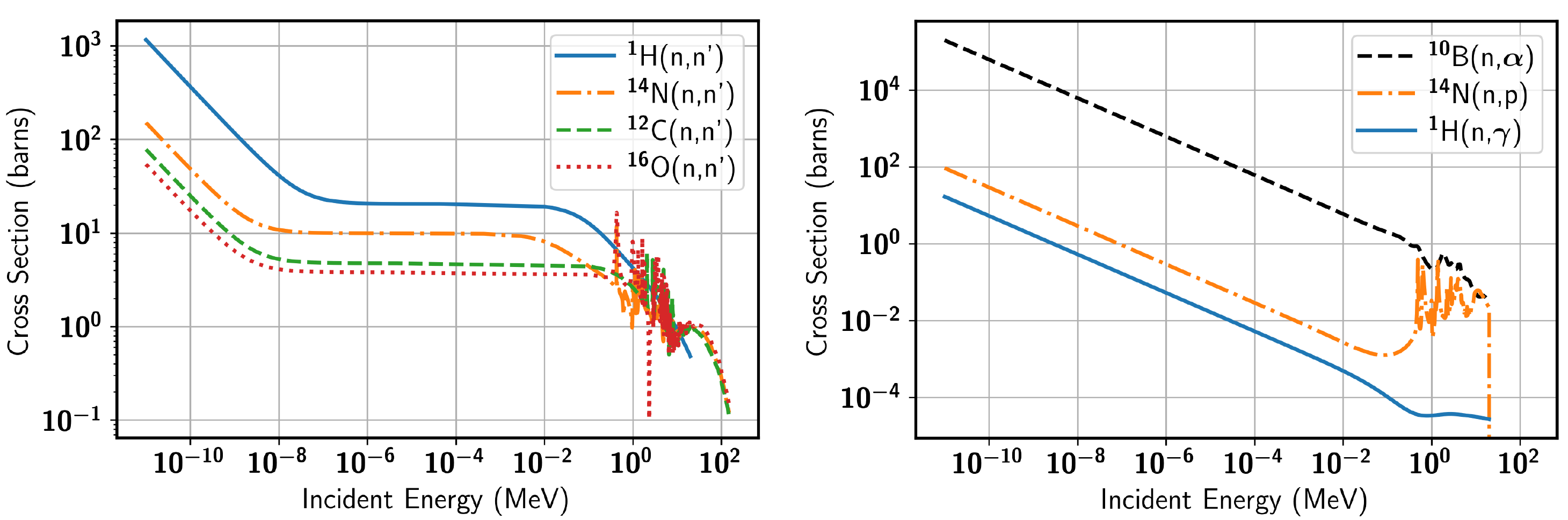}
\caption{a. (left)Elastic scattering cross section from ENDF data base for $^{1}$H, $^{14}$N, $^{12}$C and $^{16}$O ; b. (right) Neutron cross section from ENDF data base for $^{10}$B(n,$\alpha$)$^{7}$Li, $^{14}$N(n,p)$^{14}$C and $^{1}$H(n,$\gamma$)$^{2}$H }
\label{endf}
\end{figure*}

The occurrence of each nuclear reaction depends on its associated cross section, a function of the neutron energy.
Three of them are neutron capture reactions with cross section following a 1/v law when neutron energy decreases in the thermal range energy \citep{NeutronPhysics}, v being the neutron velocity,  as shown in Figure \ref{endf}. Thus, thermal neutron fields penetrate poorly into tissues and cannot reach deep seated tumours. Epithermal neutrons fields, which thermalise as they penetrate the tissue, ensure to obtain thermal neutrons at the tumour level to maximise capture probability on $^{10}$B.

Moreover, in the epithermal energy range, cross-section values for elastic neutron scattering reactions vary from less than $1\%$ for $^{12}$C and $^{16}$O to $7\%$ for $^{1}$H and up to a maximum of $19\%$ for $^{14}$N, see Figure \ref{endf}.
Neutrons with energy between 1 eV and 10 keV have a quite similar probability to interact by elastic scattering process.
As thermalisation is achieved through elastic scattering on tissue nuclei, this process seems almost independent of the neutron energy in the eptihermal range.
Then, it could be assumed that the penetrability of neutrons inside the brain does not strongly change with their initial epithermal energy as it was pointed out in \citep{Fairchild1989}. For elastic neutron scattering reactions, the energy transferred to the nuclei is proportional to the neutron incident energy. So the secondary dose induced by scattering reactions increases with higher epithermal neutron energy.
In this context, reducing the neutron energy from the present reference energy of $10$ keV to lower epithermal energies will increase the absorbed dose in the tumour while reducing the secondary dose resulting from elastic scattering reactions on healthy tissues. This hypothesis is explored hereafter by Monte Carlo simulations.
\subsection{Monte-Carlo simulations}
\subsubsection{Simulation geometry}

Simulations have been processed with the Monte Carlo code MCNP. The geometry includes a human head phantom with a brain tumour. The universe around the phantom was set empty to reduce calculation time and to ensure the initial value of the neutron energy reaching the phantom.

\begin{figure}
\centering
\includegraphics[scale=0.5]{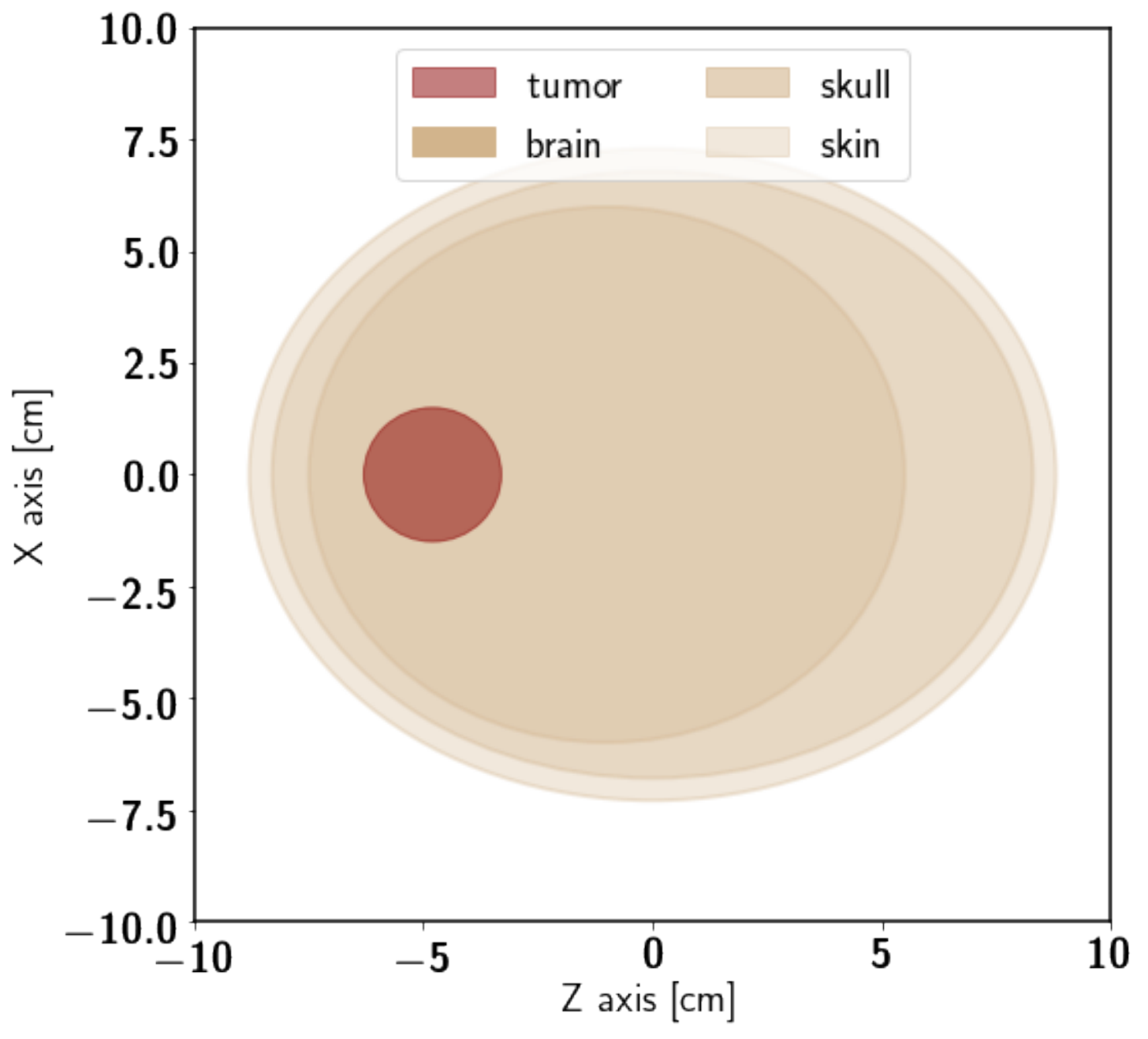}
\caption{Snyder head phantom model with tumour at 4 cm depth, ZX projection}
\label{snyder_phantom}
\end{figure}

The phantom considered here is a Snyder head model phantom \citep{Goorley2002} with a spherical tumour as shown in Figure \ref{snyder_phantom} and tissues having the composition and density recommended by ICRU-$46$\citep{ICRU46} (see Table \ref{ICRU}).

\begin{table*}[!h]
\centering
\begin{adjustbox}{max width=\textwidth}
\begin{tabular}{|c|c|c|c|c|c|c|c|c|c|c|c|c|}
  \hline
   & 
   &
  \multicolumn{11}{c|}{Weight $\%$}\\
  \hline
  tissue & $\rho$ (gr/cm$^{3}$) & H & C & N & O & Na & Mg & P & S & Cl & K & Ca  \\
  \hline
  Skull & 1.61 & 5 & 21.2 & 4 & 43.5 & 0.1 & 0.2 & 8.1 & 0.3 &  &  & 17.6\\
  Brain & 1.04 & 10.7 & 14.5 & 2.2 & 71.2 & 0.2 &  & 0.4 & 0.2 & 0.3 & 0.3 & \\
  Skin & 1.09 & 10 & 20.4 & 4.2 & 64.5 & 0.2 &  & 0.1 & 0.2 & 0.3 & 0.1 & \\
  \hline
  \hline
\end{tabular}
\end{adjustbox}
\caption{Densities and compositions for brain, skull and skin tissues, ICRU-$46$}
\label{ICRU}

\end{table*}

Surface equations for  brain, skull and skin in the Snyder model used were respectively determined by the following equations: 

\begin{equation}
(\frac{x}{6})^{2} + (\frac{y}{9})^{2} + (\frac{z-1}{6.5})^{2} = 1
\label{sur1}
\end{equation}
\begin{equation}
(\frac{x}{6.8})^{2} + (\frac{y}{9.8})^{2} + (\frac{z}{8.3})^{2} = 1
\label{sur2}
\end{equation}
\begin{equation}
(\frac{x}{7.3})^{2} + (\frac{y}{10.3})^{2} + (\frac{z}{8.8})^{2} = 1
\label{sur3}
\end{equation}

A tumour with a volume of $14.14$ cm$^{3}$, consisting of a $1.5$ cm radius sphere, is located at different depths inside the brain. The tumour is centred in X and Y and is translated on the Z axis of the geometry. Tumour depth is set between $4$ cm and $8$ cm, with $1$ cm step, to cover five non-superficial cancer cases, and having the same composition as the brain tissue. 

$^{10}$B concentrations are added to the ICRU-$46$ tissues composition, with 52.5 ppm in the tumour, 15 ppm in the brain, 22.5 ppm in the scalp and 15 ppm in the skull \citep{Capoulat2016}. The $^{10}$B concentration ratios for healthy tissues and tumours compared to $^{10}$B concentration in blood are listed in Table \ref{ratio}. The blood $^{10}$B concentration considered is $15$ ppm. 

\vspace{2mm}

\begin{table}[H]
\begin{center}
\begin{tabular}{|c||c|c|c|c|} 
 \hline
 tissue & tumour & brain & skull & skin \\ 
 \hline
 tissue to blood ratio & 3.5 & 1 & 1 & 1.5 \\ 
 \hline
\end{tabular}
\caption{Tissue to blood ratios of $^{10}$B concentration for healthy tissues and tumour.}
\label{ratio}
\end{center}
\end{table}

\begin{figure}
\centering
\includegraphics[scale=0.5]{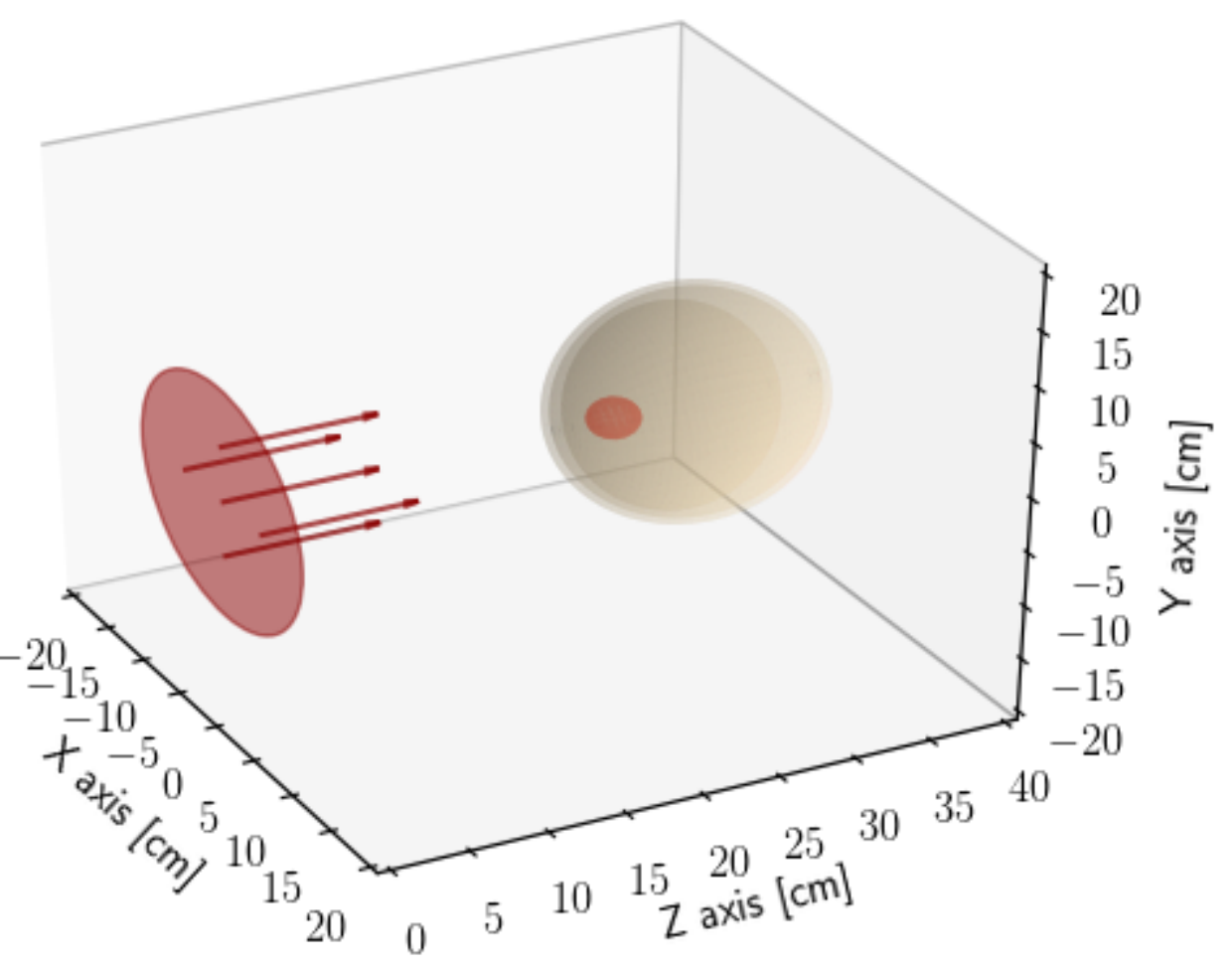}
\caption{Snyder head phantom model with a tumour at 4 cm depth, 3D view showing the neutron source location and its emission direction}
\label{snyder_phantom_+_source}
\end{figure}

The neutron source was defined as a $21$ cm diameter disc, simulating a Beam Shape Assembly (BSA) output of 10.5 cm radius and allowing to provide a complete irradiation of the head phantom. Neutrons are sent with a mono-energetic energy and uniformly on the disc along the Z axis, as it can be seen in Figure \ref{snyder_phantom_+_source}. 
In order to explore the whole epithermal range, simulations were performed at the following energies: $10$ eV, $100$ eV, $1$ keV and $10$ keV.

\subsubsection{Simulation physics model}
Due to the energy range of interest (from thermal to epithermal neutrons), thermal treatment is implemented in the simulations. Free-gas treatment is used in the simulation model above $4$ eV, representing the energy threshold below which thermal scattering laws are implemented with the use of S($\alpha$,$\beta$) functions \citep{thermalneutron}. These functions treat a specified material in the thermal regime as a molecular compound, taking into account the effects on cross sections of molecular structure and vibration modes. Only some materials are tabulated \citep{MCNPmanual}. As human tissues are similar to light water in composition and density, the tabulated data of light water were selected. 

MCNP uses variance reduction techniques, including implicit capture and weight cut-off parameters.
The weight is a value allocated to a particle that is calculated at each step of its track according to the probability of the current event to happen.
In general, the weight of the tracked particle is reduced at each interaction (scattering or capture) until a default threshold is reached below which the particle is killed.
This implicit mode, used for tally computations to obtain capture rates and total absorbed dose from a specified particle type in users specified volumes, allows computing time reduction for simulations and uncertainty calculations.

	However, this calculation method does not allow to distinguish some dose components, such as the energy deposited in ionisation by recoil protons resulting from elastic scattering reactions of neutrons with hydrogen nuclei in tissues. This dose component quantification has been determined using a PTRAC output file, which enables to obtain all user-filtered particle events \citep{MCNPmanual}. Default variance reduction techniques are turned off for the simulation, leading to the use of so-called analog capture, process during which reduction weight is not performed and neutrons are killed if capture occurs. Moreover, light and heavy ion recoil physics models have been used. These parameters provided an output file with all interactions needed to determine energy transfer between neutrons and tissue nuclei.	
	
	Two simulation sets were computed for each geometry to obtain dose components values: from tallies on the one hand to estimate capture rate for $D_{B}$ and $D_{p}$ calculation and to estimate the energy deposited by photons for $D_{\gamma}$ calculation and from PTRAC analysis, on the other hand, for estimating energy transferred by elastic scattering to hydrogen nuclei for $D_{n}$ calculation. \\
\indent One estimated standard deviation is used to compute the observable uncertainties  proper to each dose component.
	The relative uncertainties are maintained under 1 $\%$ for every dose component estimation.

\subsubsection{Simulation observables}

The neutron capture reaction rate on $^{10}$B in tissues as a function of the energy was the first quantitative observable studied, to show the impact on the boron dose $D_{B}$ \ref{sec:neutron_interactions}. Capture rates are normalised per unit mass of tissue to obtain observables proportional to the boron dose $D_{B}$. The secondary dose in healthy tissues was explored, to understand the correlation between neutron dose and proton dose variations when reducing the neutron energy.
As the main objective is to compare treatment efficacy of different mono-energies, the Therapeutic Gain (TG) was calculated. 
It is defined as the ratio between the total biological dose in the tumour and the total biological dose in the brain. 

This paper is structured as follows : first, the dose induced by the boron capture in the tumour is presented; then, a complete study of dose components in healthy tissues is described, considering a tumour at 4 cm depth in order to evaluate the impact of the neutron energy on healthy tissues dose components; the influence of the tumour depth on the total biological dose in healthy tissues is discussed afterwards; the results of the total biological dose in the tumour and in healthy tissues as a function of the neutron energy are also presented; finally, the study of the TG is shown.

\section{Results}

In the following sections, results will be presented using a relative difference to compare $1$ keV, $100$ eV and $10$ eV to $10$ keV simulated data. 
It is expressed in percentage and obtained with the following formula:
\begin{equation}
\delta = (\frac{results(E)}{results(10 keV)} - 1)*100
\end{equation}
This allows to compare results from lower epithermal energy to the results obtained at $10$ keV and to quantify the differences between values in order to conclude on the most useful energy.

\subsection{\textbf{Study of the dose in the tumour: focus on the boron dose $D_{B}$ }} \label{section_D_B_tumour}

\begin{figure*}[htpb!]
\centering
\includegraphics[scale=0.10]{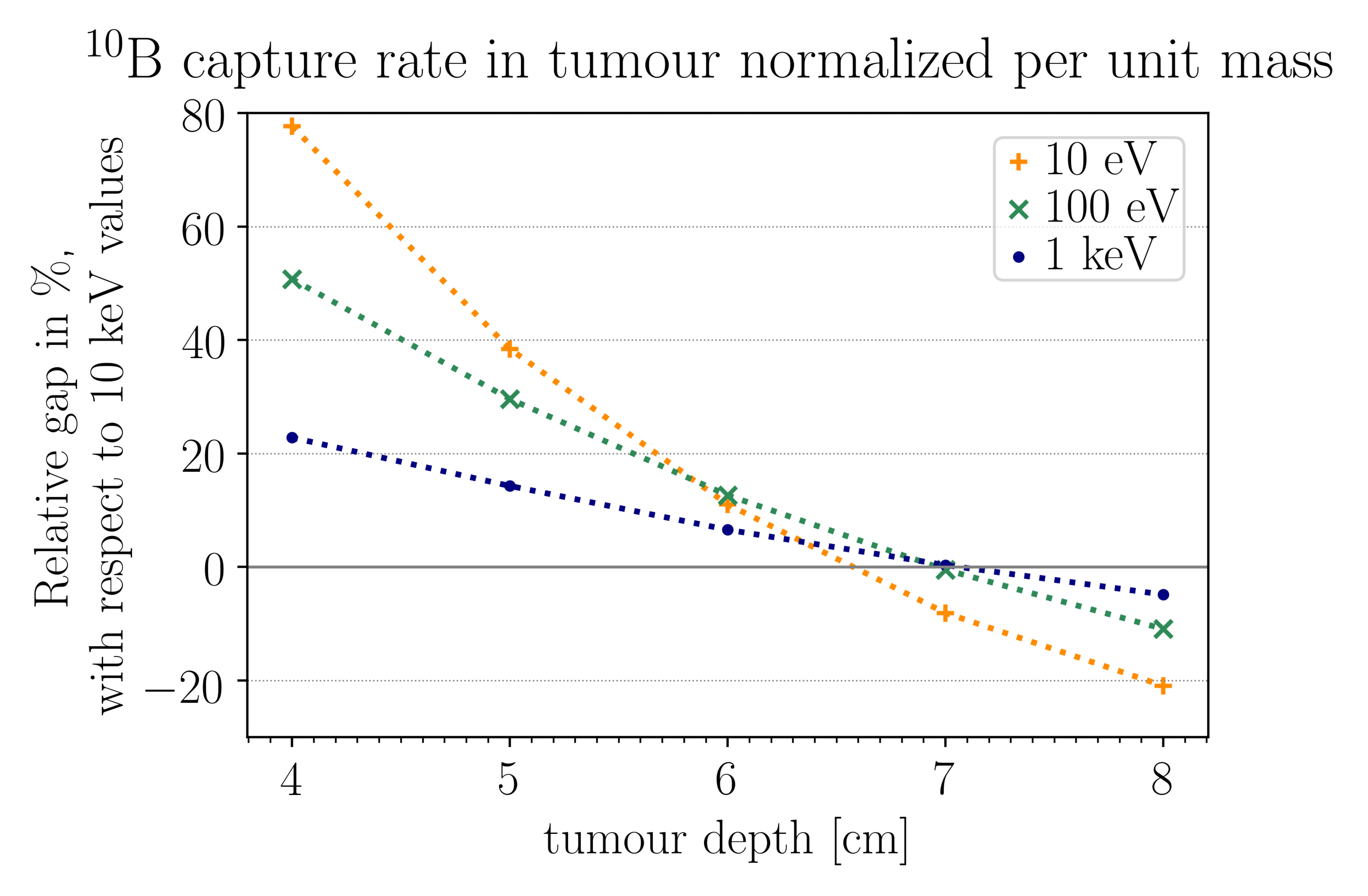}
\caption{Relative difference in percentage of neutron capture reaction rate on $^{10}$B in the tumour per mass unit, between the different epithermal energies and $10$ keV, as a function of tumour depth, with 52.5 ppm of $^{10}$B in the tumour.}
\label{boron_capture_tumour_relative_gap}
\end{figure*}

 In this section, a focus is made on the boron dose $D_{B}$ in the tumour representing the therapeutic dose for BNCT treatments. To estimate the boron dose $D_{B}$, the neutron capture reaction rate on $^{10}$B in the tumour normalized per unit mass of tissue and per neutron source is studied.
Figure \ref{boron_capture_tumour_relative_gap} shows the relative difference in percentage with respect to $10$ keV values of $1$ keV, $100$ eV and $10$ eV results.
%
For a tumour seated at 4 cm depth in tissues, the number of captures on $^{10}$B in tumour increases by 78 $\%$ with 10 eV neutrons and by 23 $\%$ with 1 keV neutrons, in comparison with capture rates obtained for 10 keV neutrons.
For deeper tumours, at 6 cm depth for example, the number of $^{10}$B captures is 11 $\%$ and 13 $\%$ higher at 10 eV and 1 keV respectively, and 13 $\%$ higher at 100 eV than at 10 keV.
Neutrons at 10 eV energies induce less captures on $^{10}$B than 10 keV neutrons for tumours located at depth higher than 6.6 cm into tissues. Neutrons at 100 eV and 1 keV induce less captures than 10 keV neutrons for tumours located at depth higher than 7 cm.

\subsection{\textbf{Secondary dose on healthy tissues: study  with the tumour at 4 cm depth}}
 
\label{sec:HT_tumour_4cm}
In the following sections, observables on healthy tissues are studied with the tumour at $4$ cm depth. The aim is to compare doses components of healthy tissues when reducing the neutron kinetic energy in the epithermal range.
The influence of the tumour depth on the total biological doses of healthy tissues will be studied in section \ref{sec:tumour_impact_HT_dose}.

\subsubsection{Neutron Capture reaction rate on \texorpdfstring{$^{10}$B}, in healthy tissues : brain, skin and skull}

In order to understand the effects of reducing neutron epithermal energy on the boron dose $D_{B}$ in healthy tissues, neutron capture reaction rates on $^{10}$B were evaluated for healthy tissues with a tumour at $4$ cm depth and normalised per unit mass. Table \ref{HT_boron_capture} presents the obtained results. 
\begin{table*}
\begin{center}
\begin{tabular}{c c c c c c c c c} 
 \hline
 & \multicolumn{2}{c}{10eV}& \multicolumn{2}{c}{100eV}& \multicolumn{2}{c}{1keV}& \multicolumn{2}{c}{10keV}\\
 & g$^{-1}$($10^{-16}$) & {\%} & g$^{-1}$($10^{-16}$) & {\%} & g$^{-1}$($10^{-16}$) & {\%} & g$^{-1}$($10^{-16}$) & {\%} \\
 \hline 
 brain & 28.23$\pm$0.02 & \textit{17} & 27.38$\pm$0.02 & \textit{13} & 25.69$\pm$0.02 & \textit{5} & 24.22$\pm$0.02 & / \\ 
 skull & 11.06$\pm$0.01 & \textit{74} & 8.98$\pm$0.01 & \textit{41} & 7.41$\pm$0.01 & \textit{17} & 6.35$\pm$0.01 & / \\
 skin & 13.56$\pm$0.01 & \textit{87} & 10.75$\pm$0.01 & \textit{48} & 8.69$\pm$0.01 & \textit{20} & 7.24$\pm$0.01 & / \\
 \hline
\end{tabular}
\caption{Absolute values and relative difference in percentage with respect to $10$ keV values of neutron capture reaction rates on $^{10}$B per mass unit on healthy tissues and per neutron source, for lower epithermal energies (for a tumour at $4$ cm depth, with 15 ppm of $^{10}B$ in brain and skull and 22.5 ppm of $^{10}B$ in skin).}
\label{HT_boron_capture}
\end{center}
\end{table*}

An important increase of $^{10}$B capture rates for the skull and the skin can be observed when considering $100$ eV or $10$ eV energy irradiation, with up to 74 $\%$ and 87 $\%$ increase respectively in comparison with 10 keV energy neutrons. For the brain, the number of neutron captures on $^{10}$B rises by $17{\%}$ at maximum when reducing the epithermal energy to $10$ eV. Relative gaps with respect to 10 keV values for the skin and the skull are higher than relative gaps with respect to 10 keV values for the brain.
Moreover in absolute values, we observed that neutron capture rates on $^{10}$B in the skull were always lower than values obtained in the skin, values even lower than in the brain, implying that the brain is the healthy tissue with the highest boron dose $D_{B}$.

\subsubsection{Study of the proton dose $D_{p}$ in healthy tissues}

 The absorbed proton dose $D_{p}$ was evaluated for healthy tissues.  
The absorbed proton dose $D_{p}$ results from ionisation of protons produced by the $^{14}$N(n,p)$^{14}$C reaction. 
Table \ref{Dp_HT} shows the relative difference in percentage with respect to $10$ keV values of results obtained for lower epithermal energies.

\begin{table*}
\begin{center}
\begin{tabular}{c c c c c c c c c}
 \hline
 & \multicolumn{2}{c}{10eV}& \multicolumn{2}{c}{100eV}& \multicolumn{2}{c}{1keV}& \multicolumn{2}{c}{10keV}\\
 & Gy($10^{-16}$) & {\%} & Gy($10^{-16}$) & {\%} & Gy($10^{-16}$) & {\%} & Gy($10^{-16}$) & {\%} \\
 \hline 
 brain & 8.44$\pm$0.01 & \textit{17} & 8.18$\pm$0.01 & \textit{13} & 7.68$\pm$0.01 & \textit{6} & 7.24$\pm$0.01 & / \\ 
 skull & 6.01$\pm$0.01 & \textit{74} & 4.88$\pm$0.01 & \textit{41} & 4.03$\pm$0.01 & \textit{17} & 3.45$\pm$0.01 & / \\
 skin & 5.16$\pm$0.01 & \textit{87} & 4.09$\pm$0.01 & \textit{48} & 3.31$\pm$0.01 & \textit{20} & 2.76$\pm$0.01 & / \\
 \hline
\end{tabular}
\caption{Absolute values in Gy and relative difference in percentage with respect to $10$ keV values of the absorbed proton dose $D_{p}$ in healthy tissues, for lower epithermal energies (for a tumour at $4$ cm depth, with 15 ppm of $^{10}B$ in brain and skull and 22.5 ppm of $^{10}B$ in skin).}
\label{Dp_HT}
\end{center}
\end{table*}

It is interesting to note that the relative gap of the absorbed proton dose $D_{p}$ in healthy tissues of energies in [10 eV; 100 eV; 1 keV] with respect to results obtained at 10 keV is similar to the one obtained considering the $^{10}$B(n,$\alpha$)$^{7}$Li reaction rate normalized per unit mass.
This can be explained by the fact that the $^{14}$N(n,p)$^{14}$C reaction used in the computation of the absorbed proton dose $D_{p}$ and the $^{10}$B(n,$\alpha$)$^{7}$Li reaction have both analogous cross section variation in the epithermal range.
The study of the absolute values shows that the brain is the healthy tissue receiving the highest value of dose from $^{14}$N(n,p)$^{14}$C reaction for all energies. 

\subsubsection{Study of the neutron dose $D_{n}$ in healthy tissues}

\begin{figure*}
\centering
\includegraphics[scale=0.15]{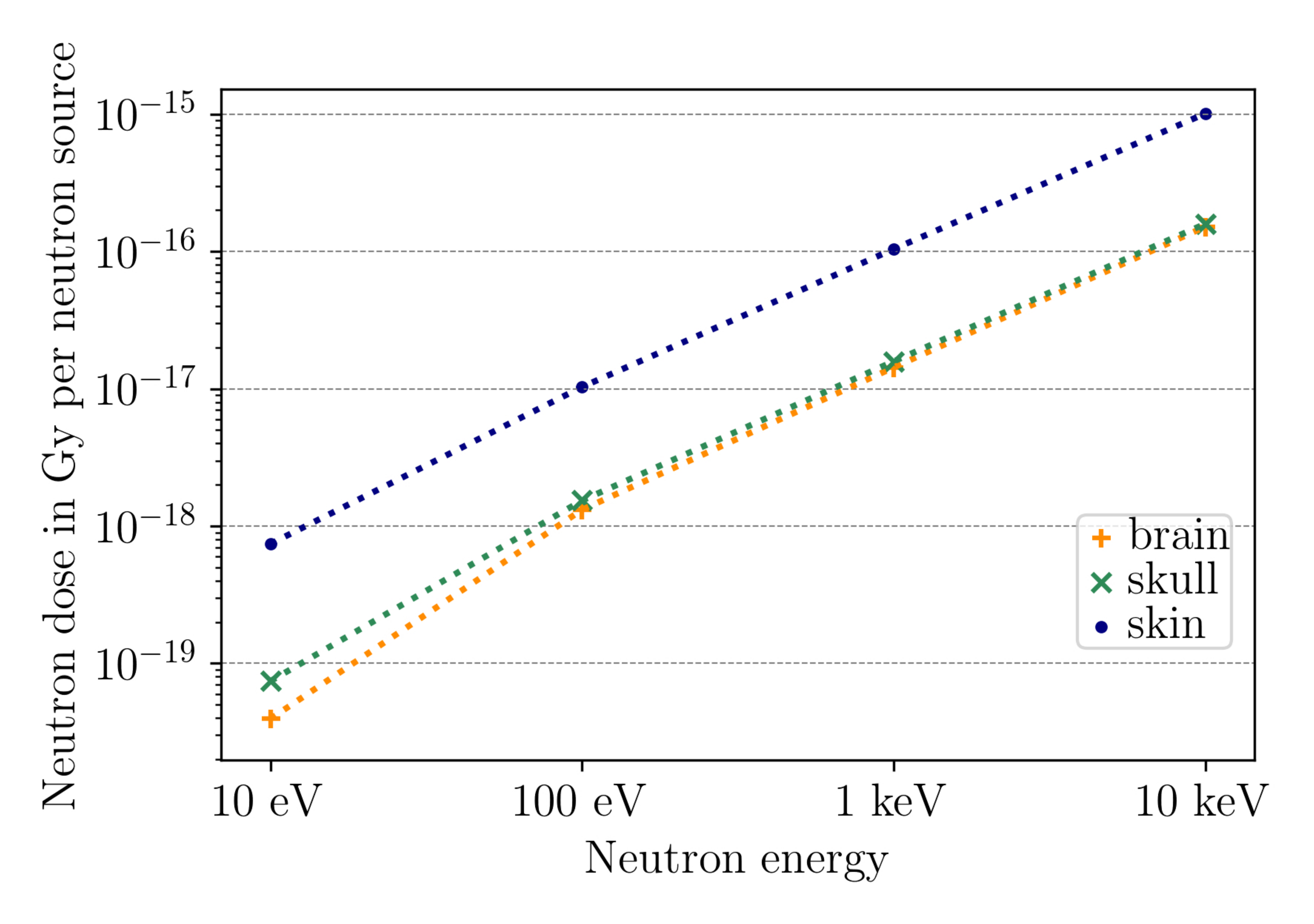}
\caption{Absorbed neutron dose $D_{n}$ in healthy tissues as a function of the neutron incident energy, expressed in Gy and per neutron, with 15 ppm of $^{10}B$ in the brain and the skull and 22.5 ppm of $^{10}B$ in the skin.}
\label{Dn_HT}
\end{figure*}

The absorbed neutron dose $D_{n}$ results from proton ionisation of protons produced by $^{1}$H(n,n')$^{1}$H' elastic scattering process. 
Figure \ref{Dn_HT} shows $D_{n}$ in the brain, the skull and the skin, evaluated for different epithermal neutron energies, still in the case of a head phantom with a spherical tumour of 3 cm radius located at 4 cm depth.

These results show that reducing the initial neutron energy by an order of magnitude enables a reduction of one order of magnitude (or more) on the value of the absorbed neutron dose $D_{n}$, on every healthy tissue. 
The variation of the neutron dose $D_{n}$ in healthy tissues is higher than the variation of the proton dose $D_{p}$ and the boron dose $D_{B}$ for the same energy range.
In this case, the skin is the tissue associated to highest values of $D_{n}$, for all energies studied.

\subsubsection{Study of the sum of the proton dose $D_{p}$ and the neutron dose $D_{n}$ in healthy tissues}

The sum of the proton dose $D_{p}$ and the neutron dose $D_{n}$ was studied to evaluate the dose contribution of proton ionisation from capture reaction. Results presented in Table \ref{sum_proton_dose} and the relative difference in percentage with respect to values obtained at $10$ keV is also indicated.

\begin{table*}
\begin{center}
\begin{tabular}{c c c c c c c c c} 
 \hline
 & \multicolumn{2}{c}{10eV}& \multicolumn{2}{c}{100eV}& \multicolumn{2}{c}{1keV}& \multicolumn{2}{c}{10keV}\\
 & Gy($10^{-16}$) & {\%} & Gy($10^{-16}$) & {\%} & Gy($10^{-16}$) & {\%} & Gy($10^{-16}$) & {\%} \\
 \hline 
 brain & 8.44$\pm$0.02 & \textit{-4} & 8.20$\pm$0.02 & \textit{-6} & 7.82$\pm$0.02 & \textit{-11} & 8.76$\pm$0.02 & / \\ 
 skull & 6.01$\pm$0.01 & \textit{19} & 4.89$\pm$0.01 & \textit{-3} & 4.18$\pm$0.01 & \textit{-17} & 5.04$\pm$0.01 & / \\
 skin & 5.17$\pm$0.01 & \textit{-60} & 4.19$\pm$0.01 & \textit{-67} & 4.35$\pm$0.01 & \textit{-66} & 12.87$\pm$0.01 & /\\
 \hline
\end{tabular}
\caption{Absolute values in Gy per neutron source and relative difference in percentage with respect to $10$ keV values of the sum of the absorbed proton dose $D_{p}$ and absorbed neutron dose $D_{n}$ in healthy tissues for lower epithermal energies (for a tumour at $4$ cm depth, with 15 ppm of $^{10}B$ in brain and skull and 22.5 ppm of $^{10}B$ in skin).}
\label{sum_proton_dose}
\end{center}
\end{table*}
 The same profile can be observed for all tissues: reducing the neutron epithermal energy from $10$ keV to lower epithermal energies appears to reduce the total dose produced by proton ionisation, as the relative gap is negative for all energies between 10 eV and 1 keV and for every healthy tissues.
The only exception is at 10 eV, where the sum of the absorbed proton dose $D_{p}$ and absorbed neutron dose $D_{n}$ in the skull is 19 $\%$ higher than at 10 keV.
It can be noted that the lowest value of the sum of the proton dose and the neutron dose is reached for $1$ keV energy neutrons. 

\subsubsection{Study of the gamma dose $D_{\gamma}$ in healthy tissues}

The absorbed gamma dose $D_{\gamma}$ results mainly , in the frame of this study, from the radiative capture reaction  $^{1}$H(n,$\gamma$)$^{2}$H on hydrogen as no photon flux is implemented in the source. 
Table \ref{photon_dose_HT} shows $D_{n}$ in the brain, the skull and the skin, evaluated for different epithermal neutron energies, still in the case of a head phantom with a spherical tumour of 3 cm radius located at 4 cm depth.

\begin{table*}
\begin{center}
\begin{tabular}{c c c c c c c c c}
 \hline
 & \multicolumn{2}{c}{10eV}& \multicolumn{2}{c}{100eV}& \multicolumn{2}{c}{1keV}& \multicolumn{2}{c}{10keV}\\
 & Gy($10^{-15}$) & {\%} & Gy($10^{-15}$) & {\%} & Gy($10^{-15}$) & {\%} & Gy($10^{-15}$) & {\%} \\
 \hline 
 brain & 8.71$\pm$0.08 & \textit{20} & 8.35$\pm$0.08 & \textit{15} & 7.77$\pm$0.07 & \textit{7} & 7.28$\pm$0.07 & / \\ 
 skull & 5.67$\pm$0.05 & \textit{28} & 5.24$\pm$0.05 & \textit{18} & 4.78$\pm$0.04 & \textit{8} & 4.44$\pm$0.04 & / \\
 skin & 4.06$\pm$0.03 & \textit{35} & 3.70$\pm$0.03 & \textit{23} & 3.32$\pm$0.03 & \textit{10} & 3.02$\pm$0.03 & / \\
 \hline
\end{tabular}
\caption{Absorbed photon dose $D_{\gamma}$ in healthy tissues as a function of the neutron incident energy, expressed in Gy and per neutron, with 15 ppm of $^{10}B$ in brain and skull and 22.5 ppm of $^{10}B$ in skin.}
\label{photon_dose_HT}
\end{center}
\end{table*}

The photon dose $D_{\gamma}$ in healthy tissues decreases when the neutron energy increases.
This variation is similar to the variation of the neutron dose $D_{n}$ and the boron dose $D_{B}$, as the cross sections associated to the reactions characterising those dose components equally vary in the epithermal range.
But the relative gap with respect to 10 keV values is not the same for the photon dose $D_{\gamma}$ and those two others dose components as the 2.2 MeV photon produced by the radiative capture on hydrogen will not be entirely contained in the phantom and will escape the geometry, releasing only a fraction of its energy in tissues.

\subsection{\textbf{Total biological dose in tissues}}

The previous section has presented the different components of the dose deposited in the healthy tissues and the $^{10}$B capture rate normalised per unit mass in the tumour. 
Hereafter, we focus on a global analysis, based on the sum of all the components and using RBE factors to obtain the total biological dose.

\subsubsection{Total biological dose in tumours}
\begin{figure*}
\centering
\includegraphics[scale=0.15]{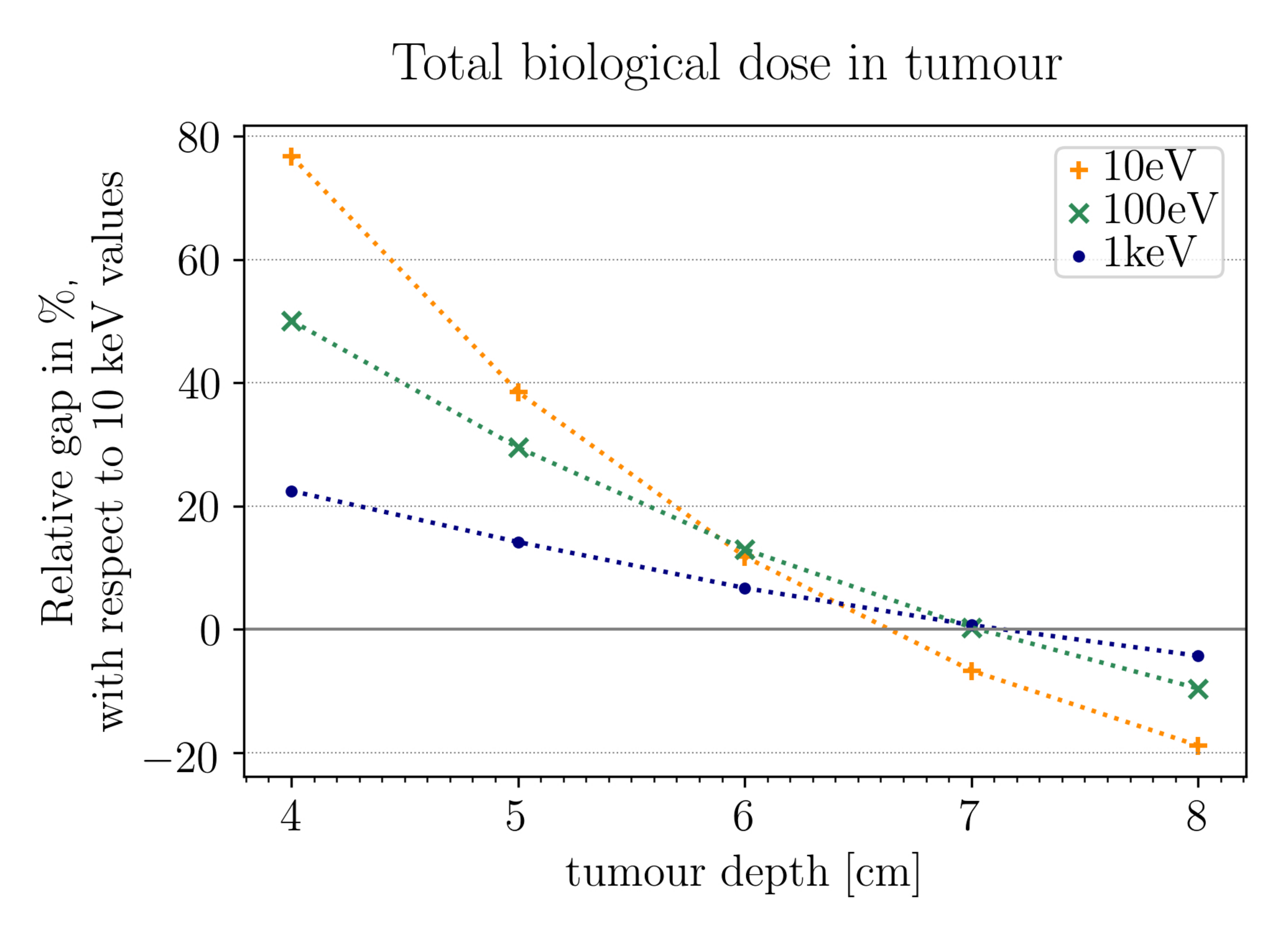}
\caption{Relative difference in percentage of the total dose in the tumour, between the different epithermal energies with respect to $10$ keV, with $52.5$ ppm of $^{10}$B in tumour}
\label{tumour_total_dose_gap}
\end{figure*}

The relative difference between the total biological dose deposited in the tumour tissue at $10$ keV and values obtained at lower epithermal energies is represented in Figure \ref{tumour_total_dose_gap}.

Total biological dose in tumour results show the same trend as the previous results obtained for $^{10}$B neutron capture reaction rate in tumour: lower epithermal energy deliver more dose inside the tumour for depth up to $6.3$ cm for $10$ eV energy, up to $6.8$ cm for $100$ eV and even up to $7$ cm for $1$ keV.
This equivalence can be explained by considering that the boron dose $D_{B}$ is the major component of the total biological dose in tumour due to the $^{10}$B concentration in the tumour.

\subsubsection{Total biological dose in healthy tissues}

The total biological dose in healthy tissues was studied in the case of a tumour at $4$ cm depth.
Table \ref{HT_total_dose} represents the total biological dose in healthy tissues.
\begin{table*}
\centering
\begin{tabular}{c c c c c c c c c} 
 \hline
 & \multicolumn{2}{c}{10eV}& \multicolumn{2}{c}{100eV}& \multicolumn{2}{c}{1keV}& \multicolumn{2}{c}{10keV}\\
 & Gy($10^{-15}$) & {\%} & Gy($10^{-15}$) & {\%} & Gy($10^{-15}$) & {\%} & Gy($10^{-15}$) & {\%} \\
 \hline 
 brain & 20.00$\pm$0.01 & \textit{15} & 19.29$\pm$0.01 & \textit{11} & 18.08$\pm$0.01 & \textit{4} & 17.45$\pm$0.01 & /\\ 
 skull & 10.95$\pm$0.01 & \textit{37} & 9.53$\pm$0.01 & \textit{19} & 8.37$\pm$0.01 & \textit{5} & 7.98$\pm$0.01 & /\\
 skin & 13.65$\pm$0.01 & \textit{20} & 11.33$\pm$0.01 & \textit{-0.4} & 9.79$\pm$0.01 & \textit{-14} & 11.37$\pm$0.01 & /\\
 \hline
\end{tabular}
\caption{Absolute values in Gy[RBE] per neutron source and relative difference in percentage with respect to $10$ keV values of the total biological dose in healthy tissues for lower epithermal energies (for a tumour at $4$ cm depth, with 15 ppm of $^{10}B$ in brain and skull and 22.5 ppm of $^{10}B$ in skin).}
\label{HT_total_dose}
\end{table*}

The total biological dose in healthy tissues increases when the incident neutron energy is lowered. 
Except for the skin, where the total biological dose decreases when reducing the energy from 10 keV to 1 keV or 100 eV. 
Our results on absolute values of the total biological dose in healthy tissues show that the brain received the higher value of dose, for every energy studied.

\subsection{\textbf{Influence of the tumour position on healthy tissue observables}}

\label{sec:tumour_impact_HT_dose}
The influence of the tumour depth was studied for every dose component of healthy tissues (being the neutron dose $D_{n}$, the proton dose $D_{p}$, the boron dose $D_{B}$ and the gamma dose $D_{\gamma}$) and for the total biological dose in healthy tissues. 
Table \ref{HT_total_dose_cm} presents the relative gap in $\%$ of the total biological dose in healthy tissues for a tumour at 4 cm depth and a tumour at 8 cm depth.

\begin{table}
\begin{center}
\begin{tabular}{c c c c c} 
 \hline
   & 10 eV & 100 eV & 1 keV & 10 keV\\ 
 \hline
 brain & 4.7 $\%$ & 4.1 $\%$ & 3.4 $\%$ & 2.7 $\%$ \\
 skull & 0.5 $\%$ & 0.5 $\%$ & 0.4 $\%$ & 0.3 $\%$\\
 skin & 0.6 $\%$ & 0.5 $\%$ & 0.5 $\%$ & 0.3 $\%$ \\
 \hline
\end{tabular}
\caption{Relative gap in percentage of the total biological dose in healthy tisses with a tumour at 8 cm depth compared to values obtained with a tumour at 4 cm, with 15 ppm of $^{10}B$ in brain and skull and 22.5 ppm of $^{10}B$ in skin.}
\label{HT_total_dose_cm}
\end{center}
\end{table}
The variation observed on the total biological dose in healthy tissues is between $0.3{\%}$ and $4.7{\%}$ for all energies and tumour depths studied.
The impact of the tumour position in the phantom can be considered as negligible, so the study made in section \ref{sec:HT_tumour_4cm} can be extended for tumours up to 8 cm depth in tissues.

\subsection{\textbf{Figure of Merit to optimise the incident neutron energy}} \label{sec:section_FOM}
The TG allows to evaluate the advantages and drawbacks of reducing neutron energies from $10$ keV to lower epithermal energies.
It is defined as following:
\begin{equation}
TG = \frac{D^{total}_{tumour}}{D^{total}_{brain}}
\end{equation}
Where D$^{total}_{tumour}$ is the total biological dose in the tumour representing the dose deposited by all the particles resulting from the neutrons interactions in the tumour and D$^{total}_{brain}$ is the total biological dose in the brain. 

In order to obtain a relevant ratio to consider the impact of the neutron energy in the epithermal range, the therapeutic ratio is evaluated considering the total biological dose in the brain.
For treatment evaluation, the maximum dose in the brain is considered for the therapeutic ratio. 
As in the frame of our study we are considering a total irradiation of the phantom, it is more relevant to consider the whole brain volume \citep{yanch1991monte}.

\begin{figure*}
\centering
\includegraphics[scale=0.16]{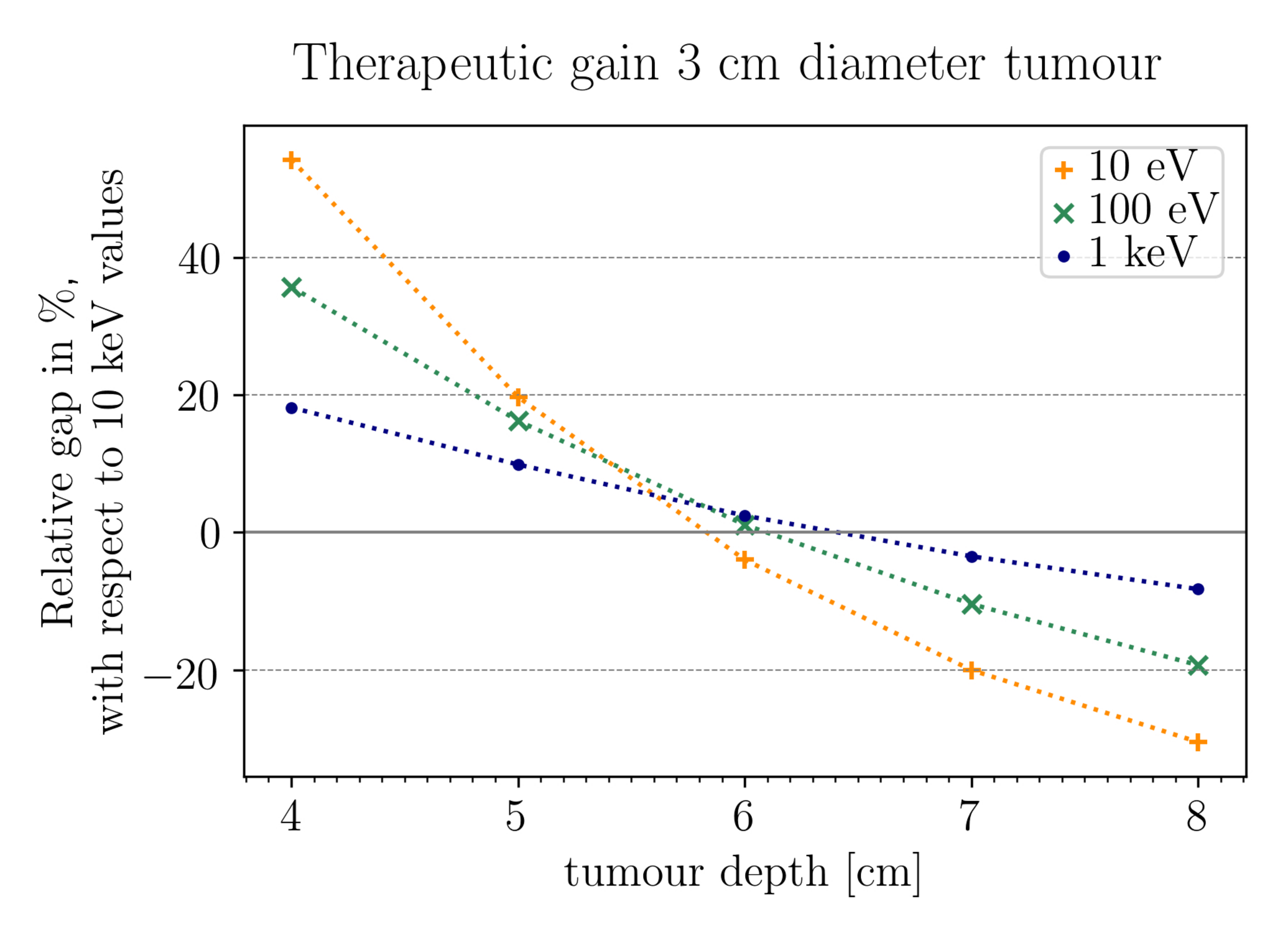}
\caption{Relative difference in percentage of ratio between total biological dose in the tumour and total biological dose in the brain, between the different epithermal energies and $10$ keV, with 15 ppm of $^{10}B$ in brain and skull and 22.5 ppm of $^{10}B$ in skin}
\label{FOM_gap}
\end{figure*}

Thus, this figure of merit allows to evaluate the optimisation of the epithermal energy.  
The relative difference in percentage between values obtained for $1$ keV, $100$ eV and $100$ eV compared with respect to $10$ keV values is shown in Figure \ref{FOM_gap}. 

For tumours seated at 4 cm depth, 10 eV, 100 eV and 1 keV neutrons provided respectively 54 $\%$, 36 $\%$ and 18 $\%$ increase on the TG compared to 10 keV neutrons. 
In the case of a 5 cm depth tumour, this increase is contained between 10 and 20 $\%$ for neutrons with energy between 10 eV and 1 keV.
The relative difference in percentage with respect to $10$ keV of lower epithermal energies studied is positive up to $5.8$ cm depth for all energies and even up to $6.4$ cm for $1$ keV energy irradiation.

\subsection{\textbf{Influence of the tumour size}}
\begin{figure*}
\centering
\includegraphics[scale=0.15]{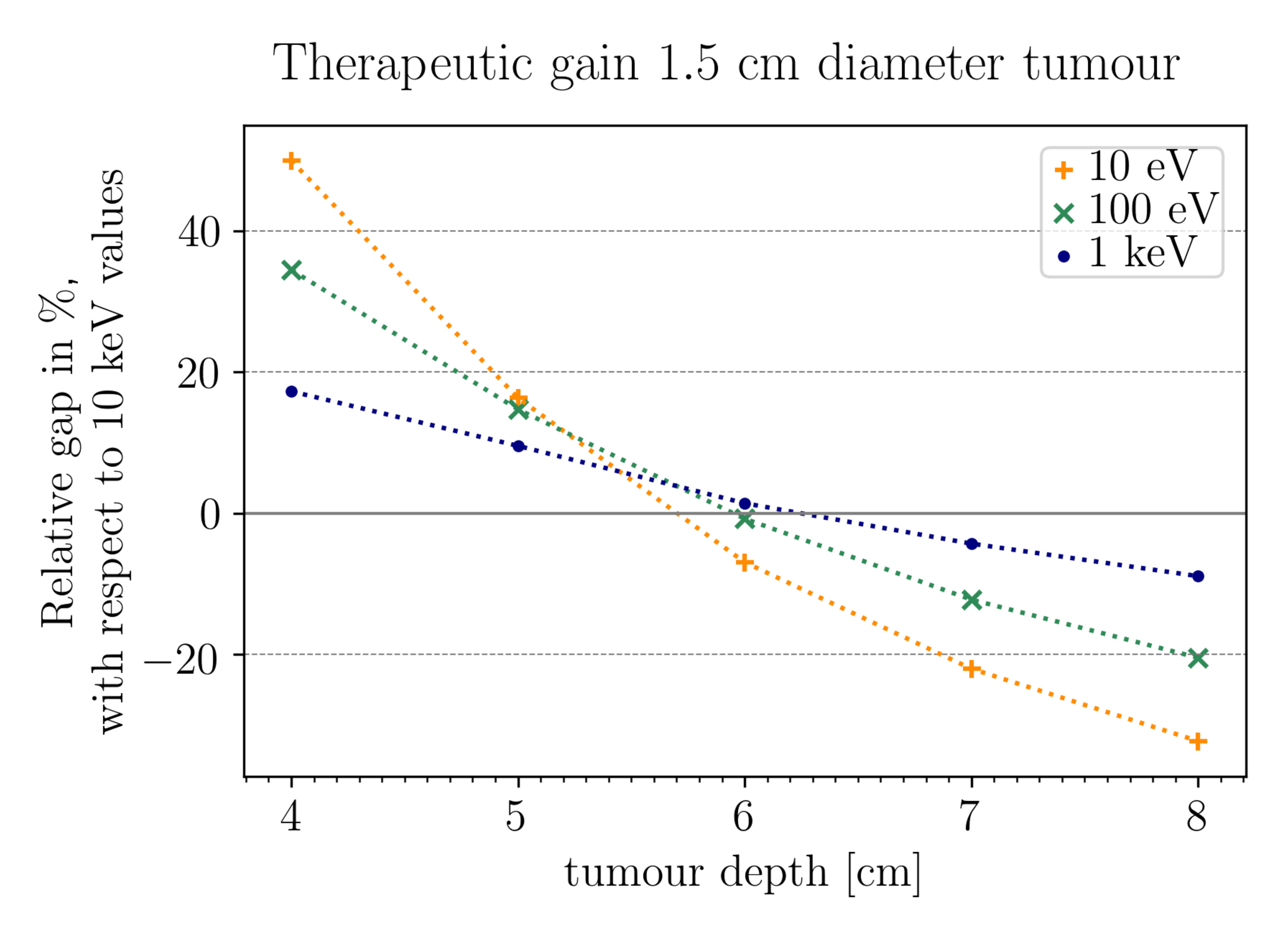}
\caption{Relative difference in percentage of ratio between total absorbed dose in tumour and total absorbed dose in brain, with a $0.75$cm diameter tumour and with 15 ppm of $^{10}B$ in brain and skull and 22.5 ppm of $^{10}B$ in skin.}
\label{tumour_1.5_gap}
\end{figure*}

In order to evaluate the influence of the tumour size on the figure of merit features, the same study was conducted with a tumour with a radius two times smaller than previously. The tumour volume was 1.77 cm$^{3}$ for a 0.75 cm radius sphere. 
Results of this study conduct to similar observations for all observables including the figure of merit proposed by this study, as pointed out in Figure \ref{tumour_1.5_gap} with the relative difference in percentage compared to $10$ keV results of the figure of merit for lower epithermal energies.

For a tumour with a volume of $1.77$ cm$^{3}$, the figure of merit reaches higher values with lower epithermal energy than $10$ keV, for tumours seated up to $5.7$ cm, and even up to $6.3$ cm for $1$ keV energy irradiation.

\section{Discussion}

Results presented in \ref{section_D_B_tumour} show that epithermal neutrons with energies lower than $10$ keV can reach deep seated tumours, up to $6.5$ cm depth, as they induce higher $^{10}$B capture rate inside the tumour. 
The results on $^{10}$B capture rate in tumours support the premise made on epithermal neutron range that penetrability does not depend strongly on their initial energy.

Nevertheless, the skull and the skin are highly impacted by the reduction of the neutron energy. Capture reaction rate on $^{14}$N and on $^{10}$B nuclei increase by 70 to 80 $\%$ when reducing from 10 keV to 10 eV the neutron kinetic energy. But the reduction of the neutron dose $D_{n}$ - dose induced by hydrogen scattering compensate for the increase of the proton dose $D_{p}$, leading to a diminution of the total absorbed dose from proton ionisation (see Table \ref{sum_proton_dose}). %
Lowering the incident neutron energy lowers the energy transferred to hydrogen nuclei : reducing the incident neutron energy reduces the dose resulting from elastic scattering of neutrons on the hydrogen nuclei in tissues.
The absorbed neutron dose $D_{n}$ declines faster than the increase of the absorbed proton dose $D_{p}$. Their sum is maximum for $10$ keV neutrons and it reaches a minimum around $1$ keV neutrons for every healthy tissue, showing the positive impact of reducing the initial epithermal energy.
If we focus on the absolute value of the total biological dose in healthy tissues, the brain is the tissue with the highest dose value for each energy studied in the epithermal range.
The use of the healthy tissue with the highest total biological dose in the definition of the TG validate the pertinence of the figure of merit.  %

As explained in \ref{sec:section_FOM}, the TG is calculated using the total biological dose in the brain as this study does not consider a realistic neutron field for a treatment but a wide, mono-energetic and unidirectional neutron field. 
The TG calculated is higher for neutrons energies below $10$ keV, for tumours located at depth up to $6.4$ cm.
	Additionally,  these results can be extended for tumours seated up to $6.5$ cm inside the head in the case of $1$ keV neutron irradiation.
	The use of epithermal energies lower than 10 keV improves the delivered dose in the tumour minimising at the same time the secondary dose in healthy tissues.
	Reducing the tumour size does not modify the main conclusion.
The maximum achievable depth for $1$ keV energy is $6.3$ cm in the case of a 1.75 cm$^{3}$ tumour, compared to $6.4$ cm calculated for the 14.14 cm$^{3}$ tumour.

The results presented are based on the computation of the biological dose. This dose is estimated using RBE factors. 
Those factors are considered as fixed and are extracted from \citep{coderre1994neutron}.
But as pointed out in \citep{gonzalez2012photon}, the use of fixes RBE factors may mislead the estimation of the biological dose in tissues. %
A more precise model is described in this paper, using experimental data of survival experiments and based on first-order repair of sub-lethal lesions and synergetic interactions between different radiations.
This model is not implemented here because it applies to specific clinical examples in the frame of realistic treatment cases. %
As the aim of this study is to compare doses induces by different epithermal energy neutron, the standard estimation of the biological dose using fixed RBE is considered as sufficient.
The results obtained in this study can be compared to those obtained in \citep{nievaart2004parameter}, where RBE factors are examined as parameters influencing the optimal neutron source energy estimation. 
Our results are in agreement with the results presented in \citep{nievaart2004parameter} : the use of neutron source in the range of 10 eV to few keV improves the capture reaction rate on $^{10}$B and the optimal neutron energy for tumours seated at 4 cm from the skin is focused around 2 keV.
Moreover, results presented in \citep{bisceglie2000optimal} indicate that neutrons with a few keV energy induce higher values of FoM, such as the TG considering the maximum biological dose in the brain. 
In the same study, the TG reaches a maximum around 3 keV for 5 cm depth tumour and for a tumour at 8 cm depth the TG is 14 $\%$ higher for 10 keV neutrons than for 1 keV neutrons (values extract from Figure 4 of \citep{bisceglie2000optimal}). In comparison, we observe that the use of 1 keV neutrons diminish the TG by 8 $\%$ compared to 10 keV neutrons (see Figure \ref{FOM_gap}) for 8 cm depth tumours. Those results can be considered as consistent as Bisceglie work is based on a different simulation parameters, such as a higher tumour to blood ratio (4.3 instead of 3.5 in this work) and lower RBE factors for protons (1.6 for $D_{n}$ and $D_{p}$ components) and $^{10}$B reaction (2.3 for $D_{B}$ component). %

The results of the simulation could be confirmed by experimental measurement of the $^{10}$B capture rate as a function of neutron kinetic energy.
This measurement could be realised, in the near future,  by the neutron spectrometer Mimac-FastN with a boron coating on the cathode used as an active phantom \citep{santos2020neutron}. 
The principle of the active phantom is based on $^{10}$B capture reaction products detection to estimate the capture rate on the tumour inside the phantom simulated by the boron coating.   
A layer of tissue equivalent material is placed between the boron coating and a well characterised mono-energetic neutron field, which energy can be set in the studied epithermal range.

\section{Conclusion}
Different mono-energetic epithermal neutron field performances were computed with MCNP and compared to the required optimized energy for AB-BNCT treatment set  at $10$ keV. The study was conducted for a tumour at different depths inside a head phantom. 

The Therapeutic Gain was defined as the ratio between the total biological dose in the tumour and the total biological dose in the brain. 
Here the whole volume of the brain is considered because the neutron field is wide and irradiate directly the totality of the phantom.
The main objective was to compare induced TG of the different mono-energetic neutron fields studied.
The relative gap between the results obtained at an eptihermal energy and the results obtained at 10 keV allow to quantify the variation in TG .
This lead to highlight the volume in the brain for which the use of epithermal energies lower than 10 keV would increase the treatment efficacy.
This volume can be defined as the layer between the surface and $6.4$ cm depth in tissues, in which the centre of the tumour can be seated.
	
	This study demonstrated that it is possible to work with epithermal energies lower than 10 keV to improve the optimisation of the delivered dose in the tumour compared to the delivered dose in the healthy tissues. As a consequence new Beam Shape Assemblies designs can be defined to adapt the neutron energy spectrum after moderation, as for a polyenergetic spectrum the effects described at each energy will add up. Moreover, as our results were provided normalized per incident neutron, the use of lower epithermal neutron energies might reduce the irradiation time needed to deliver the required dose in tumours. This irradiation time is also linked to the maximum tolerable dose to normal brain. 
	In other words, it can be assumed that the maximum exposure treatment time can be increased if higher values of dose in tumours are needed as the reduction of epithermal energies induces a reduction of the total absorbed dose in the brain, in the case of tumours seated up to $6.4$ cm in tissues. The use of $10$ keV neutrons will be more adapted for tumours seated at more than $6.4$ cm depth inside the head.
	
	Results provided by this study are promising to reduce the secondary dose in healthy tissues inside the brain but also to increase the treatment dose of AB-BNCT for deep-seated tumours and the main results might also be extended to other organs.

\section*{Acknowledgement}
We thank Olivier Meplan (LPSC) for the help provided about Monte Carlo simulations and Michael Petit ( Laboratoire de micro-irradiation, de M\'etrologie et de Dosim\'etrie des Neutrons (IRSN)) for the help on PTRAC implementation.
	
\section*{Declaration of Competing Interest}
The authors declare no competing interests.

\appendix
\section{Kerma coefficients}
\begin{figure*}
\centering
\includegraphics[scale=0.15]{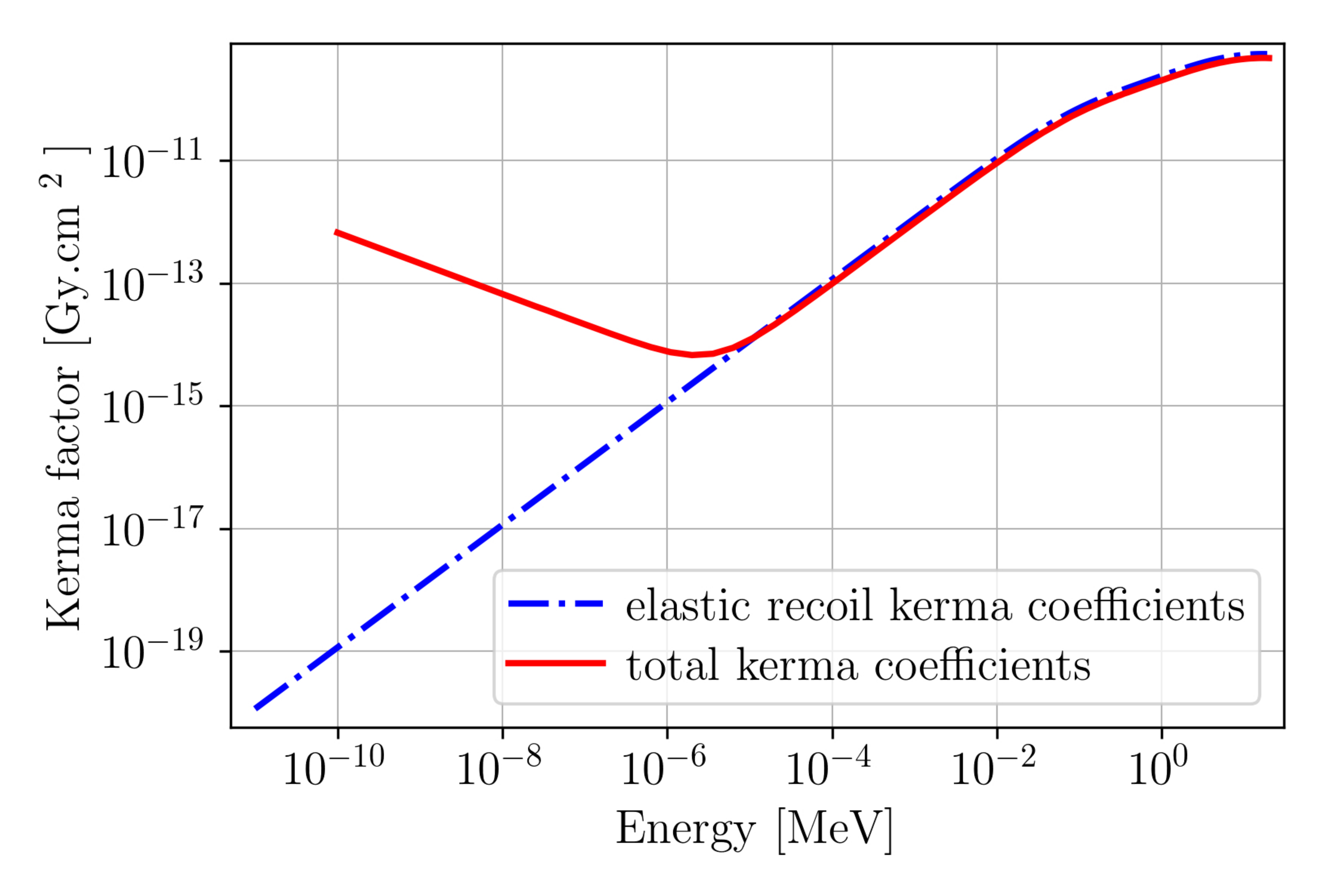}
\caption{Total kerma coefficients from \citep{Goorley2002} and calculated elastic recoil kerma coefficients for hydrogen.}
\label{kerma_1H}
\end{figure*}

\begin{figure*}
\centering
\includegraphics[scale=0.15]{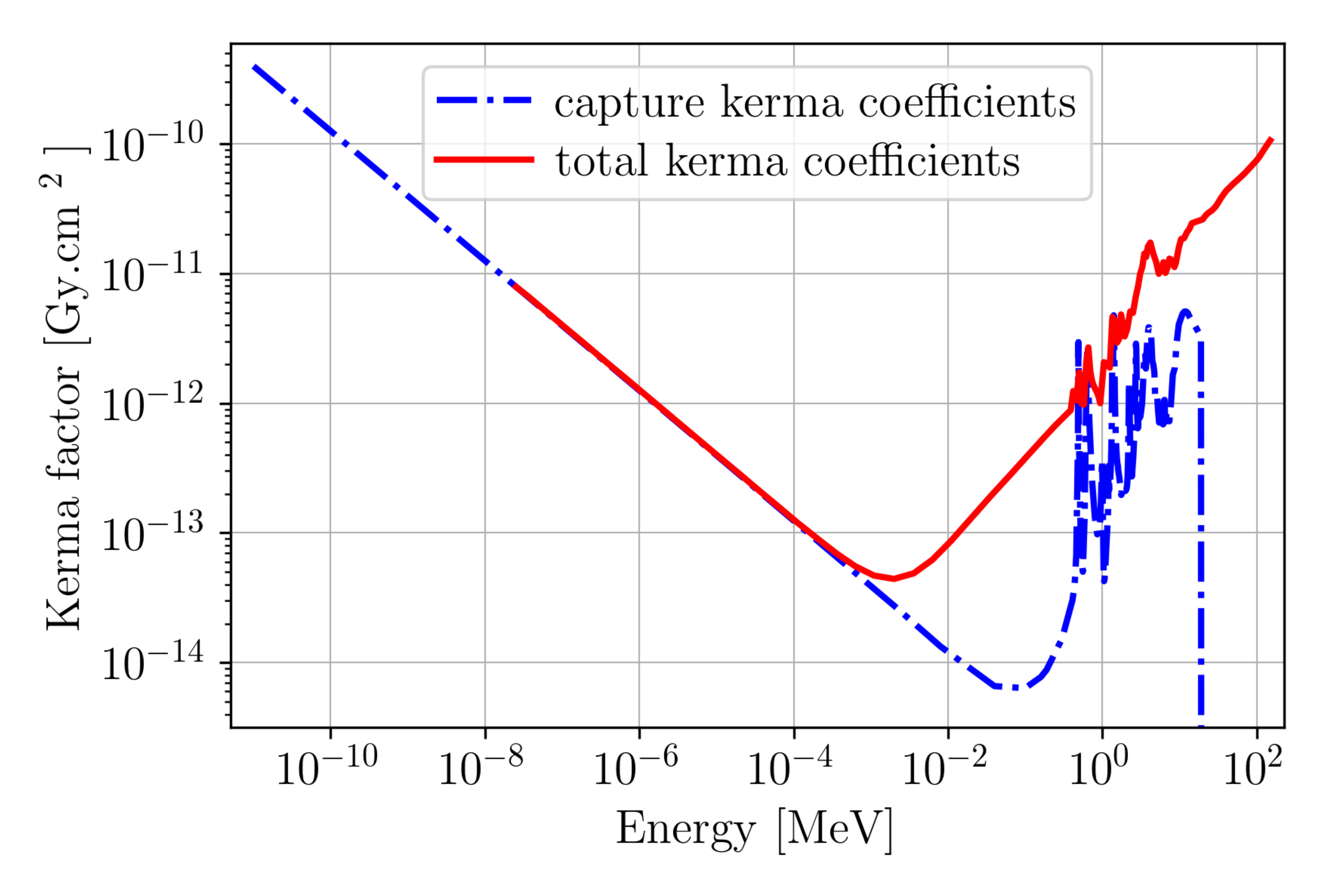}
\caption{Total kerma coefficients from \citep{Goorley2002} and calculated capture kerma coefficients for nitrogen.}
\label{kerma_14N}
\end{figure*}

Usually in dosimetry, kerma coefficients (also called kerma factors) are used to determine the dose inside a defined volume.
Kerma is an acronym for \textit{kinetic energy released in matter}. The kerma is defined as the energy transferred to charge particles per unit mass at a point of interest, including radiative-loss energy but excluding energy passed from one charged particle to another \citep{attix2008introduction} :
\begin{equation}
K = \frac{d\epsilon_{tr}}{dm}
\end{equation}
Under Charged-Particle Equilibrium (CPE), the kerma calculated in a volume is equal to the absorbed dose in the volume.
For indirectly ionizing radiations from external sources, CPE is verified in a volume if the following conditions are satisfied : the atomic composition of the medium is homogeneous; the density of the medium is homogeneous; there exists a uniform field of indirectly ionising radiation (i.e. the rays must be only negligibly attenuated by passage through the medium); no inhomogeneous electric or magnetic fields are present (chapter III section II.B \citep{attix2008introduction}).
CPE is not verified if the field of indirectly ionising radiation is not uniform in the studied volume. \\
\indent We are now considering neutrons as the indirectly ionising particles. In this case, kerma coefficients can be calculated using experimental cross section data. Total kerma coefficients are defined as the sum of kerma coefficients from each significant reaction channel of the studied element. The estimation of those coefficients requires the details of angular and energy distribution of secondary particles emitted during the reaction. 
For example, the elastic recoil kerma coefficients can be expressed as :
\begin{equation}
k_{el} = N \frac{2m_{0}M_{1}E_{n}}{{M}^2_{0}}(1-f^{m_1}_{1})\sigma_{el}
\end{equation}
where $m_{0}$ is the mass of the neutron, $M_{1}$ is the mass of the residual nucleus, $M_{0}$ is the mass of the compound nucleus, $E_{n}$ is the kinetic energy of the incident neutron, $\sigma_{el}$ is the elastic cross section and $f^{m_{1}}_{1}$ is the first Legendre coefficient of elastic angular distribution \citep{Sun2008}. This first Legendre coefficient represents the cosine of the mean angle between the recoil and the direction of the incident neutron.
Moreover, kerma coefficients induced by capture can be expressed as :
\begin{equation}
k_{capt} = N_{A} \sigma \frac{(Q_{i}+E_{n})}{M_{0}}
\end{equation}
where $\sigma$ is the capture cross section of the considered reaction, $Q_{i}$ is the Q value of the reaction and $E_{n}$ is the kinetic energy of the incident neutron \citep{Goorley2002}.
Estimated total kerma coefficients and elastic recoil coefficients for hydrogen are presented in Figure \ref{kerma_1H}.
For neutron energy higher than few eV, elastic recoil kerma coefficients correspond to total kerma coefficients. 
For epithermal and fast neutron, elastic scattering is the main process inducing dose deposit in tissues when considering neutron interactions with hydrogen. 
Estimated total kerma coefficients and capture coefficients for nitrogen are presented in Figure \ref{kerma_14N}.
Here, capture kerma coefficients match total kerma coefficients up to few keV.
For neutrons in thermal range and a part of epithermal range, dose deposition is mainly achieved trough capture, regarding neutron interactions with nitrogen.
The characterisation of the neutron dose $D_{n}$ by the elastic scattering on hydrogen and the characterisation of the proton dose $D_{p}$ with the (n,p) capture on nitrogen is in agreement with kerma coefficients computation as detailed above. \\
\indent However, in the case of our study CPE is not verified, as the neutron field in the phantom is not homogeneous.
Transient Charged Particle Equilibrium (TCPE) is considered easier to achieve than CPE and allows to establish a relation of proportionality between dose and kerma.
Although, longitudinal and lateral equilibria conditions are required to achieve TCPE in the whole volume of study, as explained in \citep{Papanikolaou2004} for a photon irradiation case. 
The half-width of the radiation field must exceed the maximum lateral motion of source particle and the position of interest must be deeper than the maximum range of charged
particles. 
Due to the geometry of our problem, it cannot be assumed that in the whole phantom volume TCPE is achieved. \\
\indent Thus, for our study, kerma coefficients cannot be used to compute the total dose in tissues as CPE and TCPE are not achieved in the considered volumes of our study.
Dose computation using kerma coefficients can be done if smaller volumes are considered inside a wider one (with homogeneous composition and density and other CPE conditions are verified).
The peak dose (defined as the dose inside a 1 cm$^{3}$ volume \citep{Coderre_dose}) is calculated using kerma coefficients. The total dose in a tissue can also be calculated using kerma coefficients. For example, the total dose in the brain is estimated by summing doses inside all 1 cm $^{3}$ volumes that constitute the brain phantom modelled in the treatment planning MacNCTPlan software (used in the Harvard-MIT clinical trial of neutron capture therapy) \citep{Palmer2002}.
For the reasons detailed above the use of kerma coefficients was not considered in our paper describing and implementing an original method more adapted to be applied for large tumours inside the brain. \\



\end{document}